\documentclass[10pt, conference, letterpaper]{IEEEtran}
\IEEEoverridecommandlockouts
\usepackage{cite}
\usepackage{amsmath}
\usepackage{amsfonts}
\usepackage{algorithmic}
\usepackage{graphicx}
\usepackage{textcomp}
\usepackage{xcolor}
\usepackage{subfiles}
\usepackage{bm}
\usepackage{subfigure}
\usepackage{multirow}
\def\BibTeX{{\rm B\kern-.05em{\sc i\kern-.025em b}\kern-.08em
    T\kern-.1667em\lower.7ex\hbox{E}\kern-.125emX}}
\begin{document}

\title{Privacy-Preserving Gesture Tracking System Utilizing Frequency-Hopping RFID Signals}
\author{Bojun Zhang
\thanks{Bojun Zhang is the corresponding author. Tianjin University, China Tianjin
300000, China. Email: ewwllvraier@126.com.}}
\maketitle


\begin{abstract}
Gesture tracking technology provides users with a "hands-free" interactive experience without the need to hold or touch devices. However, current gesture tracking research has primarily focused on tracking accuracy while neglecting issues of user privacy protection and security. This study aims to develop a gesture tracking system based on frequency-hopping RFID signals that effectively protects user privacy without compromising tracking efficiency and accuracy. By introducing frequency-hopping technology, we have designed a mechanism that prevents potential eavesdroppers from obtaining raw RFID signals, thereby enhancing the system's privacy protection capabilities. The system architecture includes the collection of RFID signals, data processing, signal recovery, and gesture tracking. Experimental results show that our method significantly improves privacy protection levels while maintaining real-time and accuracy. This research not only provides a new perspective for the field of gesture tracking but also offers valuable insights for the use of RFID technology in privacy-sensitive applications.
\end{abstract}

\section{INTRODUCTION}
\subsection{motivation}
\begin{figure}[htbp]
    \centering
    \includegraphics[width=0.48\textwidth]{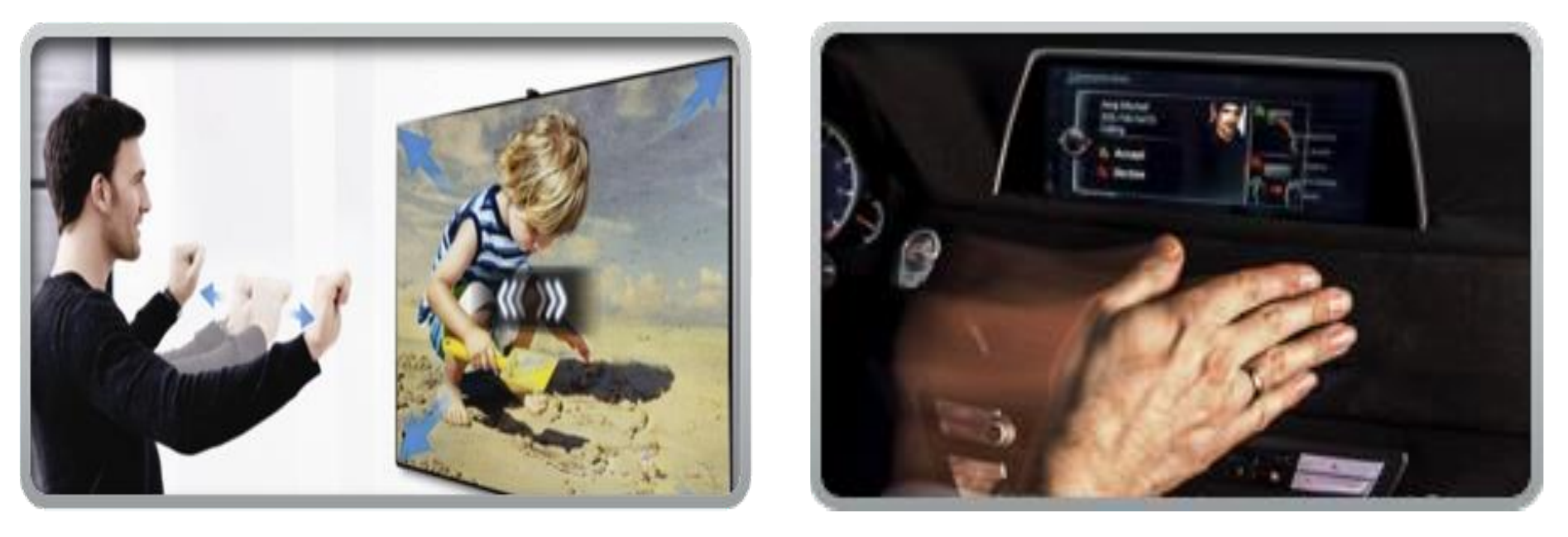}
    \caption{Gesture Recognition Schematic}
    \label{fig:enter-label0}
\end{figure}

Gesture tracking technology greatly simplifies the interaction between the user and the device, allowing the user to realize the interactive experience of “freeing their hands” without holding or tapping the device.With the advancement of technology, especially in the Internet of Things (IoT) and smart home, gesture recognition has become one of the key technologies to enhance the user experience. 
As shown in the figure \ref{fig:enter-label0}, for example, when using a tablet, users can scroll or swipe the screen with simple gestures for more intuitive and convenient operation. Similarly, when watching TV, users can change the channel and adjust the volume directly through gestures without relying on the remote control, and this non-contact interaction provides great convenience for users. In addition, with the development of large-screen and high-resolution display technology, users often need more flexible control when watching videos or browsing photos. The ability to navigate content through simple gestures such as waving a hand not only improves ease of operation, but also makes the user experience richer and more intuitive.
\subsection{Related Work}
Gesture tracking technology has garnered widespread attention due to its potential to enhance user interaction experiences. However, existing gesture tracking technologies have primarily focused on tracking accuracy while neglecting privacy protection and security issues. This section provides a review of the current mainstream gesture tracking technologies and analyzes their shortcomings in privacy protection.

\subsubsection{Computer Vision (CV)}
Computer vision technology achieves gesture tracking by analyzing video data captured from cameras. 

MediaPipe\cite{zhang2020mediapipehandsondevicerealtime} is a framework for real-time gesture tracking that runs locally on a mobile device without sending data to a remote server. This work utilizes machine learning models and optimized image processing algorithms for low latency and high accuracy gesture recognition.
Zhou et al.\cite{9354538} proposed a Spatial-Temporal Multi-Cue Network (STMC) for sign language recognition and translation tasks. They utilized a framework that integrates a Spatial Multi-Cue (SMC) module and a Temporal Multi-Cue (TMC) module, extracting multi-cue information such as hand shapes, facial expressions, and body postures from videos through a self-contained pose estimation branch. By employing a segmented attention mechanism and a joint optimization strategy.

Although computer vision has made certain advancements in gesture recognition, it requires processing large amounts of data, which not only affects the efficiency of real-time tracking but also may lead to privacy leaks due to improper handling during data storage and transmission.

\subsubsection{WiFi Tracking}
WiFi technology achieves gesture tracking by analyzing changes in wireless signals. 

Gao et al.\cite{gao2021towards} investigated a position-independent gesture recognition method using WiFi signals for gesture tracking. They proposed a novel WiFi-based gesture recognition system capable of efficiently and accurately recognizing gestures across various positions and environments.
Wu et al.\cite{wu2020fingerdraw} developed a technology named FingerDraw, which utilizes WiFi signals for sub-wavelength level finger motion tracking. Their research demonstrates the potential of WiFi signals in high-precision finger motion capture, offering new possibilities for gesture interaction on mobile devices and wearables.

However, this technology is typically less accurate and struggles to meet the demands of high-precision gesture recognition. Additionally, due to the openness of WiFi signals, there are also privacy concerns.

\subsubsection{Inertial Measurement Unit (IMU)}
IMUs track gestures by integrating accelerometers and gyroscopes. 

Cao et al.\cite{cao2021itracku} presented a system named ITrackU for tracking a pen-like instrument using Ultra-Wideband (UWB) and Inertial Measurement Unit (IMU) fusion. 
Zhang et al.\cite{zhang2021finegrainedrealtime} presented a method that leverages the high temporal resolution of IMU sensors to track and recognize a wide variety of intricate hand gestures with minimal latency.

Although IMUs can provide relatively accurate gesture tracking in certain scenarios, they are complex in design, costly, and difficult to widely deploy in real-life situations.

\subsubsection{Radio-Frequency Identification (RFID) Technology}
In recent years, RFID technology has been widely applied in the field of gesture tracking due to its low cost, ease of deployment, and high efficiency. RFID technology provides an effective technical means for gesture control by identifying and tracking subtle changes in user gestures. The advantages of using RFID for tracking include completeness, accuracy, speed, and cost-effectiveness. The tracking process is divided into two parts: determining the initial position and trajectory tracking, where the accuracy of the initial position is crucial for the entire tracking process.

RF-pen\cite{wang2020rfpen} utilizes radio frequency modulation and demodulation technology to allow users to interact with the device through gestures without actually touching the screen by creating a virtual touch interface over the air.
RF-IDraw\cite{wang2014rfidraw} creates a virtual touch screen using radio frequency signals. Users can operate it through gestures in the air without any physical screen. This technology is able to accurately track the position of the user's fingers by analyzing the reflected back RF signals, enabling touch and gesture control in the air.

Despite the widespread application of gesture tracking technology, most current work focuses solely on tracking itself without considering privacy protection and security issues. As the application of RFID gesture tracking technology becomes more widespread, its security and privacy protection issues have become increasingly prominent. Traditional RFID systems are susceptible to eavesdropping and attacks, and user gesture data may be illegally obtained and misused, which not only threatens user privacy and security but also limits the broader application of RFID technology. Therefore, this study aims to explore a gesture tracking system based on frequency-hopping RFID signals that can effectively protect user privacy while ensuring efficient and accurate tracking. The research goal is to develop an RFID gesture tracking technology that can prevent eavesdropping.
\subsection{Challenges}
RFID tracking technology, due to its efficiency and cost-effectiveness, has been widely applied in various fields. However, RFID systems are susceptible to eavesdropping when tracking objects or gestures, leading to potential privacy breaches. This security vulnerability not only threatens individual privacy but can also be exploited for malicious purposes. Therefore, ensuring user privacy without compromising the efficiency and accuracy of the tracking system presents a significant challenge.

The main challenge addressed in this study is how to achieve privacy protection for RFID signals without sacrificing the efficiency and accuracy of the tracking. To tackle this challenge, we propose a solution that involves the use of frequency-hopping signals to disrupt potential eavesdroppers, preventing them from obtaining the original RFID signals. Frequency-hopping technology increases the difficulty of illegal interception and signal decoding by constantly changing the signal frequency.

Furthermore, we utilize deep learning techniques to train a generative network model. This model is capable of recovering the original signals from the frequency-hopped signals and can be used for subsequent gesture tracking and classification tasks. In this way, we not only enhance the system's privacy protection capabilities but also maintain the accuracy and real-time performance of the tracking.
\section{SYSTEM DESIGN}
This section describes the design of our system modules.
\subsection{Principles}
\begin{figure}[htbp]
    \centering
    \includegraphics[width=0.48\textwidth]{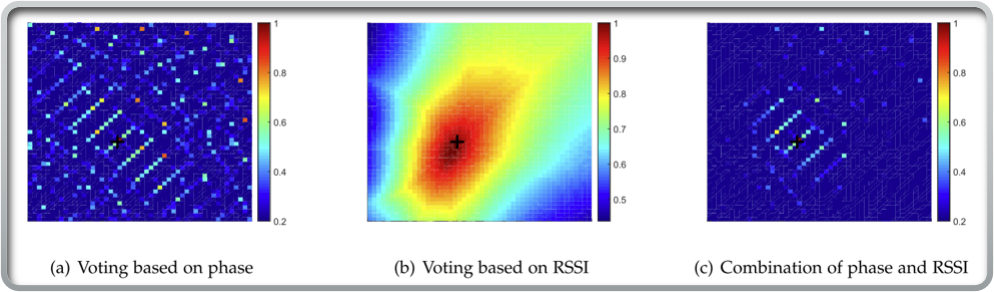}
    \caption{Schematic}
    \label{fig:enter-label1}
\end{figure}
As shown in the figure \ref{fig:enter-label1}, the principle of our gesture tracking system involves two main stages: determining the initial position and trajectory tracking. Initially, for each potential starting position of a tag, we calculate the theoretical phase difference between that position and each antenna. A normal distribution coefficient is then added to the phase difference for each antenna, and these values are summed to determine the confidence level of this initial position, denoted as \( V \). The position with the highest \( V \) value among all possible starting positions is selected as the optimal initial position\cite{2017Gyro}.

After the initial position is determined, trajectory tracking is performed using the phase differences collected by multiple antennas. This approach allows for precise tracking of the tag's movement through the analysis of the phase differences, which are indicative of the tag's position relative to the antennas.
\subsection{System Overview}
\begin{figure}[htbp]
    \centering
    \includegraphics[width=0.48\textwidth]{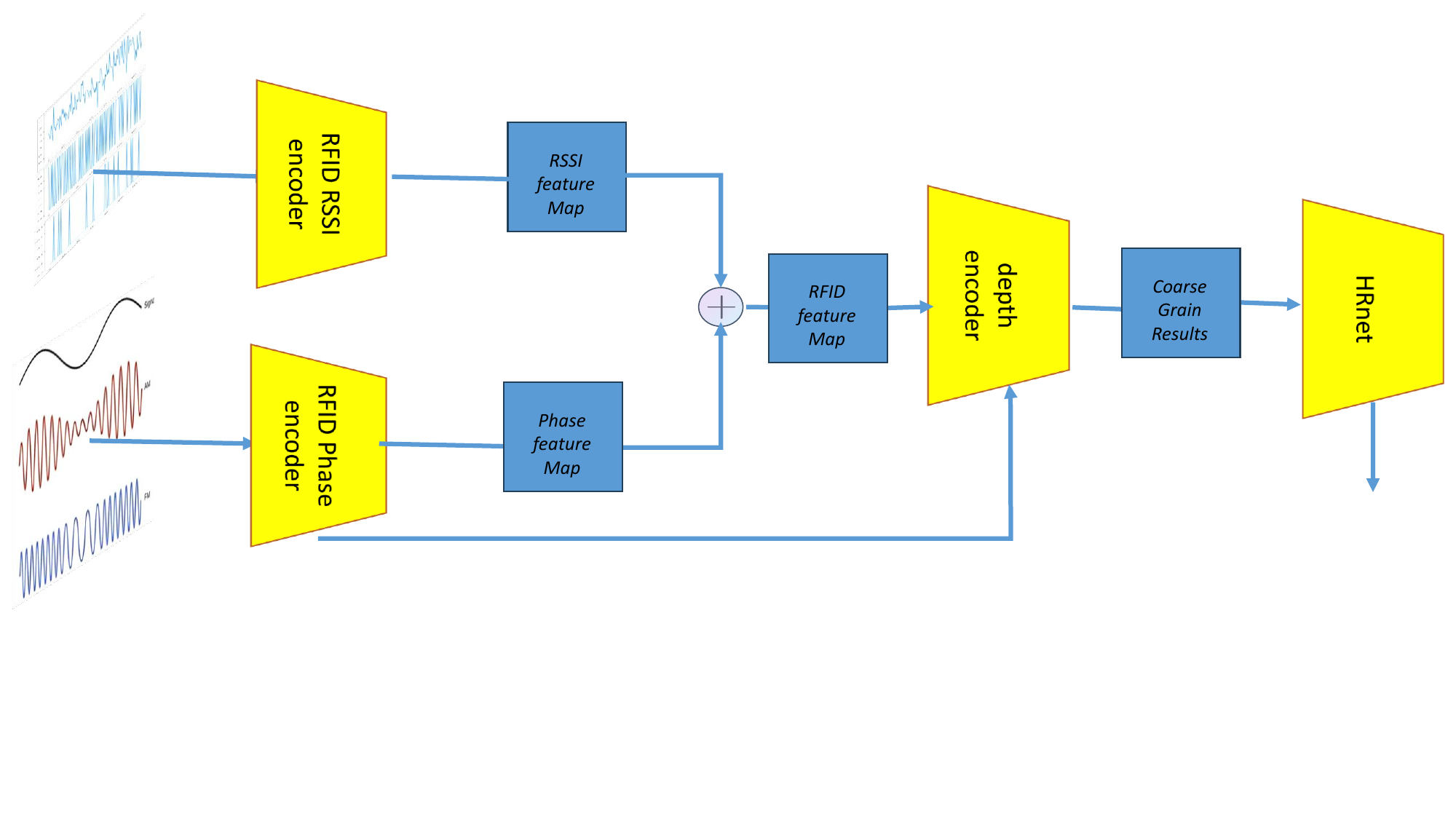}
    \caption{System Architecture}
    \label{fig:enter-label2}
\end{figure}
As shown in the figure \ref{fig:enter-label2}, this paper proposes a gesture tracking system based on frequency-hopping RFID signals, which achieves high-precision gesture tracking by collecting RFID frequency-hopping signals and combining them with Received Signal Strength Indicator (RSSI) for data processing. The inclusion of RSSI is not only because it provides vital information about the distance of the signal source but also because it encompasses environmental reflections and scattering information, which are crucial for predicting the precise location of the tag. In our system, RSSI is utilized to assist in phase prediction and uses the de-hopped RSSI signals as auxiliary labels for prediction and auxiliary supervision. This approach enables the model to learn not only the fundamental characteristics of the signals but also to prevent overfitting through auxiliary information, thereby enhancing the model's generalization capabilities. Our goal is to train a deep learning model that can recover the original signals from complex frequency-hopping signals and use these signals for effective gesture tracking and classification.
\subsection{Model Design}
\begin{figure}[htbp]
    \centering
    \includegraphics[width=0.48\textwidth]{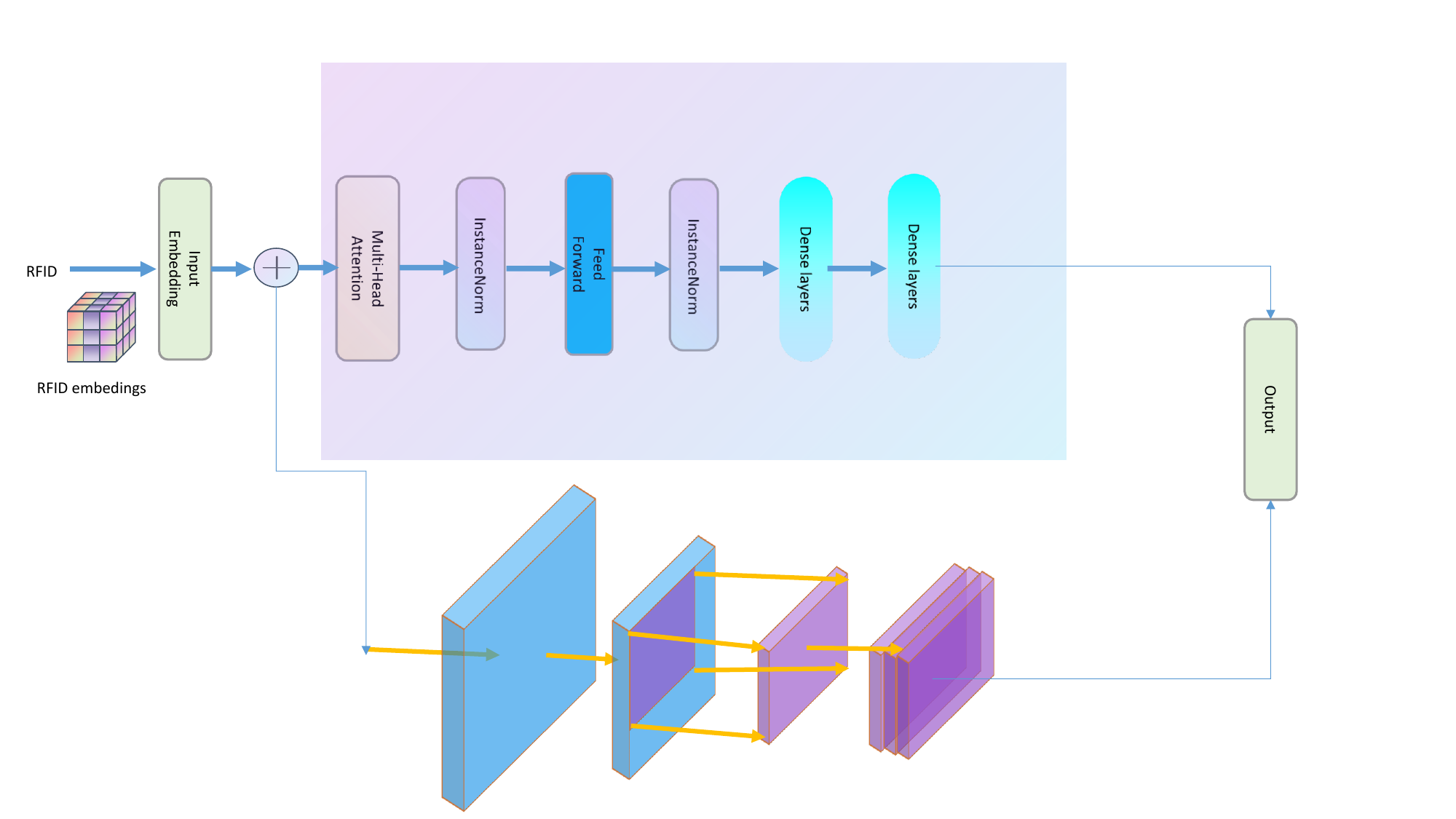}
    \caption{Model Structure}
    \label{fig:enter-label3}
\end{figure}
The structure of our model is shown in figure\ref{fig:enter-label3}, 
In this paper, we adopt the Conformer\cite{gulati2020conformerconvolutionaugmentedtransformerspeech} architecture to construct an RFID signal generation model. Conformer is an advanced neural network architecture that effectively captures local features and global dependencies by combining the strengths of Convolutional Neural Networks (CNNs) and Transformers. The application of this architecture in RFID signal processing significantly improves the precision of signal parsing and the generalization capability of the model.

The core of Conformer lies in the organic combination of its Multi-Head Self-Attention (MHSA) module and Convolution module. The Multi-Head Self-Attention module processes long-range dependencies in sequence data through the following formula:
\[
\text{Attention}(Q, K, V) = \text{softmax}\left(\frac{QK^T}{\sqrt{d_k}}\right)V
\]
where \(Q\), \(K\), and \(V\) represent the query, key, and value matrices, respectively, and \(d_k\) is the dimension of the key. This operation allows the model to focus on any part of the sequence that is relevant to the signal. The Convolution module is responsible for extracting local features from the signal, implemented through one-dimensional depthwise convolution:
\[
\text{Conv}(x) = \sum_{k} \text{ReLU}(\text{Conv1D}(x, k))
\]
where \(x\) is the input signal and \(k\) is the convolution kernel, with ReLU being the activation function. This structure enables the model to capture local details of the signal.

Furthermore, Conformer processes features through a Feed Forward Network (FFN), which consists of two linear transformations and a non-linear activation function, as shown in the formula below:
\[
\text{FFN}(x) = \max(0, xW_1 + b_1)W_2 + b_2
\]
where \(W_1, W_2, b_1, b_2\) are model parameters. This structure enhances the model's ability to express non-linear features.

The application of the Conformer architecture in the RFID generation model allows the model not only to handle complex signal variations but also to improve computational efficiency while maintaining high accuracy. This is mainly due to Conformer's dual advantages in capturing local features and global dependencies, making the model more flexible and efficient in processing RFID signals.
\begin{table*}[t]
\centering
\begin{tabular}{|c|c|c|c|c|c|c|c|}
\hline
\multicolumn{2}{|c|}{\multirow{2}*{Model}}& \multicolumn{3}{|c|}{dataset1} &\multicolumn{3}{|c|}{dataset2}\\
\cline{3-8}
\multicolumn{2}{|c|}{} & MAE & RMSE & R2  & MAE & RMSE & R2\\
\hline
\multicolumn{2}{|c|}{Ours} 	&5.30	&11.00	&0.76		&5.00	&11.60	&0.86	\\ 
\hline
\multirow{2}{*}{cnn layers}
& one layers  &7.50 & 13.81 &	0.71  &	6.96 &	13.12 &	0.73  \\
\cline{2-8}
& three layers 	&9.84	&15.30	&0.47	&8.86	&14.89	&0.53	\\
\hline
\multirow{2}{*}{cnn width}
& one layers  &6.56 & 12.52 &	0.74  &	6.47 &	12.27 &	0.79  \\
\cline{2-8}
& four layers 	&9.50	&16.02	&0.39 &8.93	&15.03	&0.46\\
\hline
\multirow{2}{*}{Transformer layers}
& zero layers  &9.68 &	16.31 &	0.53  &8.77 & 15.82 &	0.59  \\
\cline{2-8}
& two layers 	&7.48	&13.68	&0.67	&7.22	&13.47	&0.68	\\
\hline
\multirow{2}{*}{preprocessing}
& no CNN 	&7.63	&13.79	&0.69	&6.61 &12.99	&0.79	\\
\cline{2-8}
& no Transformer 	&9.44	&15.39	&0.52	&8.82	&14.82	&0.61	\\
\hline
\end{tabular}
\caption{Performance comparison of ablation experiments}
\label{table:PerformanceComparison}
\end{table*}
\section{IMPLEMENTATION}
In this section we need to introduce the experimental part.
\subsection{Experimental Setup}
\begin{figure}[htbp]
    \centering
    \includegraphics[width=0.48\textwidth]{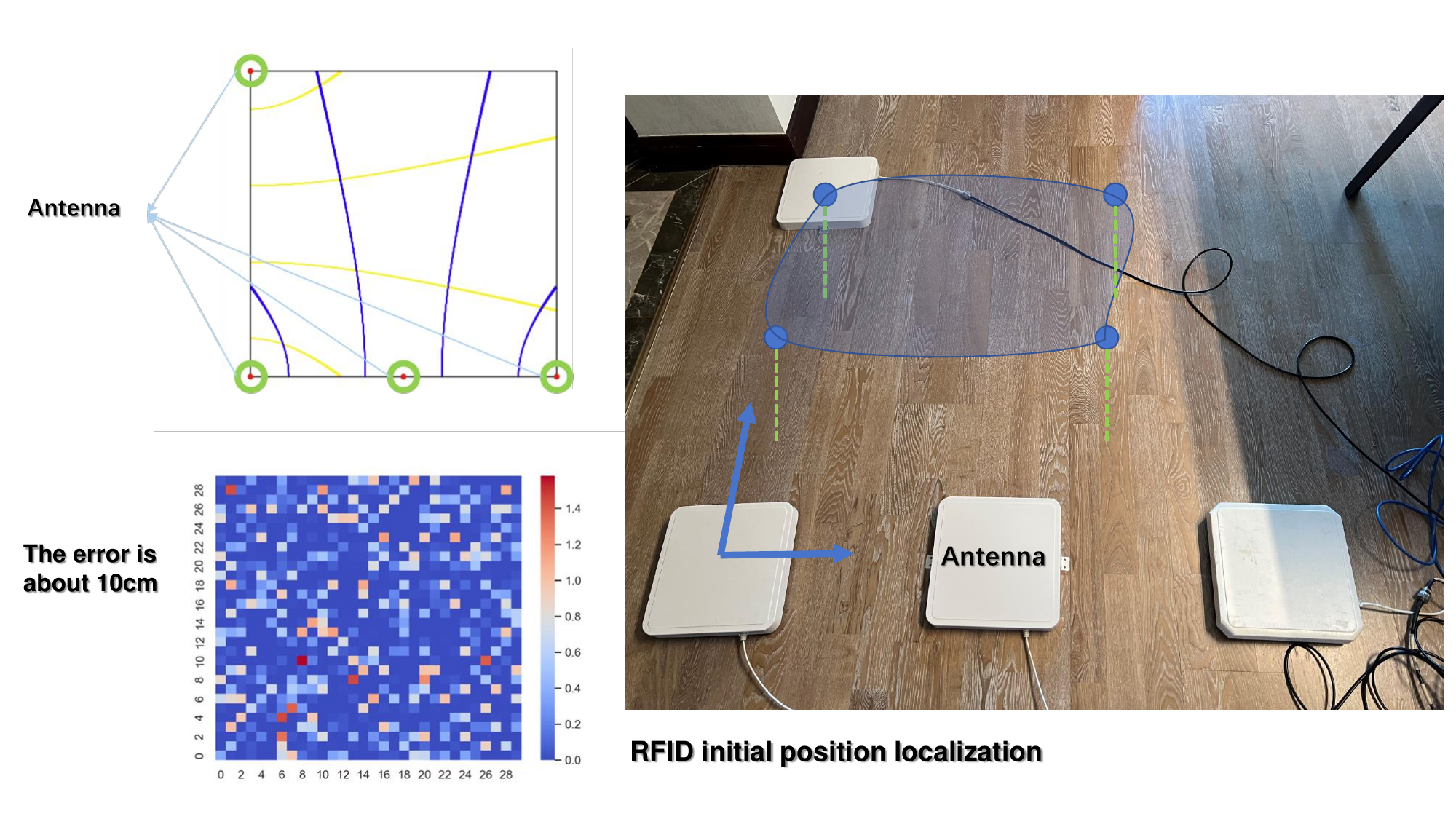}
    \caption{Experimental Scenario Setup}
    \label{fig:enter-label4}
\end{figure}
\subsubsection{Hardware}
\begin{figure}[htbp]
    \centering
    \includegraphics[width=0.48\textwidth]{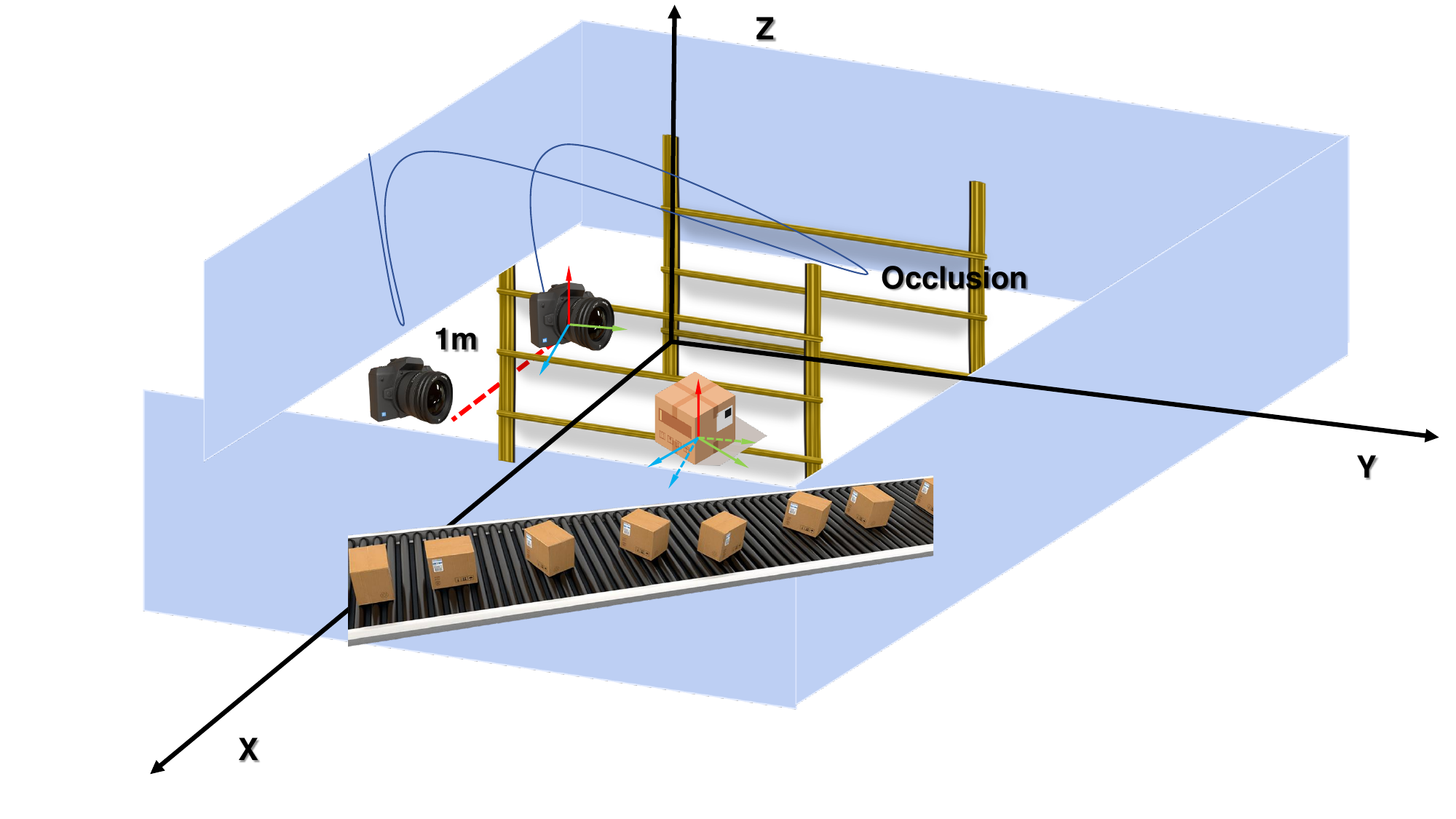}
    \caption{Schematic diagram of the experimental scene}
    \label{fig:enter-label10}
\end{figure}

As shown in the figure \ref{fig:enter-label4} and \ref{fig:enter-label10}, in this experiment, we constructed a testing environment for our RFID signal generation model based on the Conformer architecture. The experiment utilized RFID readers of the Impinj R1000 model, which are favored for their efficient signal reception capabilities and stability. To simulate the behavior of RFID tags in practical applications, we employed RFID tags based on the Alien Higgs 3 chip, programmed to send predefined signals at specific frequencies. For receiving and processing these RFID signals, we used the National Instruments (NI) USRP-2943R Software Defined Radio (SDR), equipped with two antennas, a standard configuration for RFID systems. The distance between the antennas is set at 14.0 cm, less than half the wavelength of the RFID signal to ensure effective signal reception. Additionally, we controlled the SDR with a Raspberry Pi 4, using Python scripts to capture signal samples at a sampling rate of 2MHz. In the experiment, the error for the initial position localization of the RFID is approximately 10 cm, providing crucial reference data for subsequent signal processing and model training. All collected samples are processed on a local computer equipped with an Intel(R) Core(TM) i9-11900KF @ 3.50 GHz CPU. To accelerate the training process, we also used an NVIDIA GEFORCE RTX 3080 Ti graphics card. 
\subsubsection{Software}
In our experimental process, we utilized Windows 10 Professional as the operating system to ensure stability and compatibility for our experiments. The primary programming language employed was Python 3.8, chosen for its rich library support and concise syntax. For deep learning frameworks, we used TensorFlow 2.4 and PyTorch 1.7 to build and train our models. For data processing and analysis, we relied on the powerful libraries Pandas and NumPy. Traditional machine learning algorithms and model evaluations were performed using Scikit-learn 0.24. Additionally, we employed Matplotlib and Seaborn for data visualization and Raspberry Pi 4 for controlling our RFID hardware devices. The entire experiment's coding, debugging, and version control were carried out in the PyCharm 2020.2 integrated development environment. The selection of these software and libraries ensures the efficiency and reliability of our experimental results.
\subsubsection{Training Setup}
In the model training process, we adopted the following key hyperparameters: the learning rate was set to 0.001, the batch size to 256, and the optimizer chosen was Adam, which automatically adjusts the learning rate and is suitable for large-scale parameter updates. To reduce overfitting, dropout layers with a rate of 0.1 were integrated into the model. The training process was configured to stop after 300 epochs or when the model's loss fell below a threshold of 0.01. The loss function utilized was the mean squared error (MSE), which calculates the L2 loss between the true gesture positions and the predicted ones. Weight initialization was performed using the Xavier method to maintain stable variance of activations and gradients. The nonlinear activation function in the model was ReLU, and L2 regularization with a coefficient of 0.0001 was introduced to further suppress overfitting. The careful selection and adjustment of these hyperparameters were aimed at ensuring the efficiency of the model training and the generalization performance in the end.
\begin{figure*}[htbp]
\centering
\subfigure[CDF curve of the radial distance error.]{\includegraphics[width=0.3\textwidth]{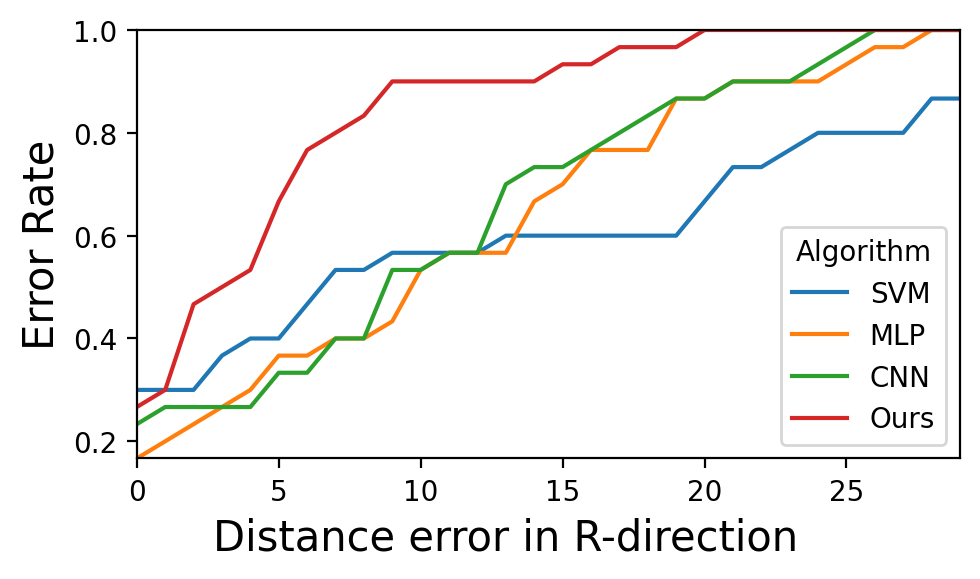}\label{fig:opa}}
\subfigure[CDF curve of the distance error in the X-direction.]{\includegraphics[width=0.3\textwidth]{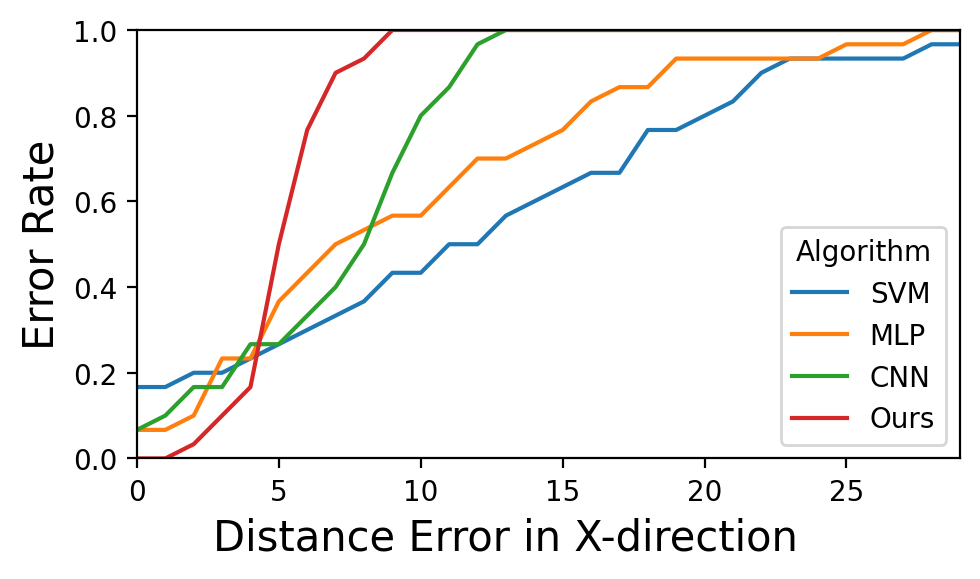}\label{fig:cvg}}
\subfigure[CDF curve of the distance error in the Y-direction.]{\includegraphics[width=0.3\textwidth]{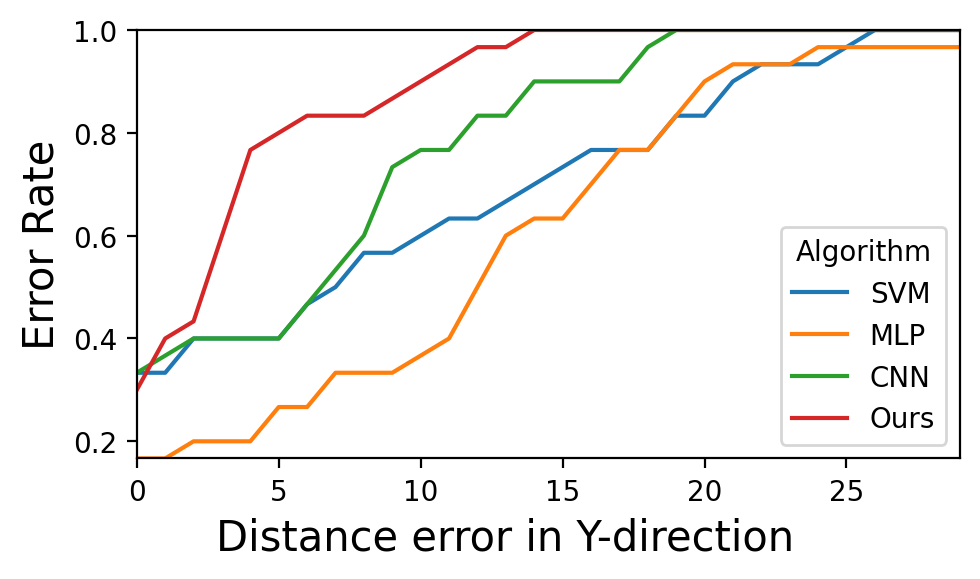}\label{fig:los}}
\caption{CDF curves in different directions.}
\label{fig:cdf}
\end{figure*}
\subsection{Experimental Evaluation Criteria}
In our control experiments, we focus on the continuous tracking of gesture keypoints in two-dimensional space. To assess the performance of the models, we employ the method of statistical absolute error. Specifically, we calculate the Euclidean distance between the predicted and true positions of the keypoints, given by $R = \sqrt{(x_{\text{pred}} - x_{\text{true}})^2 + (y_{\text{pred}} - y_{\text{true}})^2}$, where $(x_{\text{pred}}, y_{\text{pred}})$ represents the coordinates predicted by the model, and $(x_{\text{true}}, y_{\text{true}})$ represents the actual coordinates. We use the Cumulative Distribution Function (CDF) curve to display the performance of different methods in terms of error distribution.

\subsection{Baseline Methods}
To validate the effectiveness of our approach, we compare it with the following three baseline methods:
\begin{itemize}
  \item \textbf{SVM: } Support Vector Machine\cite{s23177324} is a classical machine learning method used for classification and regression tasks. In gesture tracking, SVM can be utilized to predict the positions of keypoints.
  \item \textbf{MLP: } Multiple Hypothesis Tracking\cite{liang2022mutual} is an algorithm used for dealing with multi-target tracking problems, and it can estimate the positions of keypoints in gesture tracking.
  \item \textbf{CNN: } Convolutional Neural Networks\cite{liu2023improved} are deep learning methods widely used in image processing and computer vision tasks, including gesture tracking.
\end{itemize}

By comparing with these baseline methods, we can more comprehensively evaluate the performance and advantages of our model.
\subsection{Comparison Experiment Analysis}
In this section, we compared the performance of our gesture tracking system with three baseline methods—Support Vector Machine (SVM), Multilayer Perceptron (MLP), and Convolutional Neural Network (CNN)—when tracking the X, Y, and radial (R) directions of gesture keypoints. Through Cumulative Distribution Function (CDF) curves, we demonstrated the error rate comparison of different methods in each direction. As shown in Figure \ref{fig:cdf}, our method exhibited lower error rates in all directions, indicating its superiority in tracking accuracy. Specifically, in the X-direction, our method maintained lower error rates across all distance error thresholds, showing better performance than SVM and MLP, and also outperformed CNN in the low error range. In the Y-direction, our method similarly performed excellently, with error rates consistently at a low level, further validating the system's effectiveness in tracking the Y-coordinate. CNN showed better performance in the medium error range but experienced a decline in performance as the error increased. In terms of radial distance error, our method achieved the lowest error rates across all distance error thresholds, demonstrating a significant advantage when tracking the overall position of gesture keypoints by integrating errors from the X and Y directions. In contrast, the other methods showed higher error rates, especially when the distance error was large. In summary, our gesture tracking system demonstrated superior tracking performance and lower error rates in the X, Y, and R directions, proving the effectiveness and accuracy of our method.
\subsection{Analysis of ablation experiments}
In order to verify the impact of each module in the model on the performance and to analyze the robustness of the model to different scenarios, we conducted ablation experiments. We constructed two datasets with different scenarios to verify the robustness of our model to different scenarios. We did this by replacing and removing some of our modules. The evaluation is done by MAE and RMSE and R2 coefficients.

As shown in the table \ref{table:PerformanceComparison}, the experimental results show that reducing the CNN layer leads to an increase in MAE and RMSE and a decrease in R2 coefficient, indicating that the CNN layer plays an important role in feature extraction. Among them, the performance of 1-layer CNN is slightly better than 3-layer CNN, which indicates that the width of CNN should not be too large, otherwise it will lead to model overfitting. Adding Transformer layer can improve the model performance, which indicates that Transformer has an advantage in capturing sequence features. 2-layer Transformer has better performance than 0-layer Transformer, which indicates that at least one layer of Transformer is needed to extract sequence features. Removing either the CNN or Transformer preprocessing module results in performance degradation, indicating that the preprocessing module is critical to model performance. the CNN preprocessing module effectively extracts local features, while the Transformer preprocessing module captures global dependencies, which together provide richer feature information for the model. Compared to the original model, all the ablation models show varying degrees of performance degradation on both dataset 1 and dataset 2, which further proves the necessity of the individual modules in the model. The original model achieved the best performance on both datasets, indicating the effectiveness and robustness of the model. Therefore, the results of the ablation experiments show that the CNN layer and the Transformer layer are crucial to the model performance, and the preprocessing module can effectively improve the model performance. 

Our model shows good robustness in different scenarios and provides new ideas for privacy-preserving gesture tracking techniques based on RFID signals.
\section{Conclusion}
In this paper, a gesture tracking system based on frequency hopping RFID signals is proposed, which realizes efficient and accurate gesture tracking and effectively protects user privacy by introducing frequency hopping technology and deep learning model. The results of ablation experiments show that the CNN layer and Transformer layer are crucial to the model performance, and the preprocessing module can effectively improve the model performance. Our model shows good robustness in different scenarios and provides new ideas for privacy-preserving gesture tracking techniques based on RFID signals. In the future, we will further explore more advanced signal processing techniques and try to apply the system in more practical scenarios, such as smart home and human-computer interaction.

\bibliographystyle{IEEEtran}
\bibliography{2}
\end{document}